\documentclass[12pt,a4paper]{article}
\usepackage{multirow}
\usepackage{graphicx}
\usepackage[left=2cm,right=2cm,top=2cm,bottom=2cm]{geometry}
\usepackage{amsmath}
\usepackage{color,soul}
\usepackage{xcolor}
\begin{document}

\title{RMS and charge radii in a potential model}
\author{Tapashi Das$^{1,\ast}$, D K Choudhury$^{1,2}$ and  K K Pathak$^{3}$ \\  $^1$Department of Physics,\ Gauhati University,
Guwahati-781 014, India.
\\ $^2$ Physics Academy of North-East, Guwahati-781014, India. \\
$^3$ Department of Physics, Arya Vidyapeeth College, Guwahati-781016, India\\
 $^\ast$\small{email:t4tapashi@gmail.com}}
\date{}
\maketitle

\begin{abstract}
The Dalgarno's method of perturbation is used to solve the Schrodinger's equation with the Cornell potential $V(r)=-\frac{4\alpha_s}{3r}+br+c$. The short range and long range effect of the potential is incorporated in the same wave function by using two scales $r^{S}$ and  $r^{L}$ as an integration limit. With the wave function then the results for bounds on r.m.s. radii of various heavy flavored mesons are reported. We have also showed the relation between r.m.s. and charge radius of mesons.
\end{abstract}


\section{Introduction}

Quantum Chromodynamics (QCD) \cite{1} is established to be the successful theory of strong interaction and the remaining problem in our understanding of QCD has to do with quark confinement in hadrons. It is expected that confinement is the dominant quark-antiquark or quark-quark interaction at large separation distances and therefore plays the vital role to explain the properties of highly excited (large sized) hadrons  \cite{2}. In the last few years, the heavy-heavy mesons ($\eta_c, J/\psi, \chi_c, h_c, \eta_b, \chi_b (1,2,3P), h_b, \gamma$ etc.) have been widely studied theoretically as well as experimentally. Again highly excited light-quark mesons show a hydrogen-like spectral pattern  \cite{3} and can be reproduced within a non relativistic constituent quark model framework which becomes asymptotically coulombic  \cite{4}. Thus the distance or scale behaviour of the QCD potential in studying the static and dynamic properties of hadrons is still a topic to explore. The method of perturbation in this context has the advantage of choosing parent (more dominant) and perturbative(less dominant) term from the QCD potential. The advantage of taking Cornell Potential for study is that it leads naturally to two choices of “parent” Hamiltonian, one based on the Coulomb part and the other on the linear term, which can be usefully compared  \cite{5}. Aitchison and Dudek in ref. \cite{6} showed that with Coulombic part as parent (VIPT), bottomonium spectra are well explained than charmonium, where as charmonium states are well explained with linear part as parent.\\

In this spirit a QCD potential model has been pursued for the two choices in seperation in ref. \cite{7,8,9} by considering the Dalgarno's method  \cite{10,11} of perturbation. The results include static and dynamical properties of heavy flavored mesons, like their form factors, masses, decay constants as well as Isgur wise function. Here in this work, we emphasise in the ground state of heavy-light mesons as well as $c\bar{c}$ and $b\bar{b}$ states, though $c\bar{c}$ and $b\bar{b}$ has numbers of excited states for different `$l$' and `$s$' values.\\

However, while using perturbative method, one should consider two aspects of Quantum Mechanics: (a) The scale factor `c' in the potential should not affect the wave function of the meson even while using perturbation theory to be compatible with quantum mechanical idea and\\
(b) The specific choice of perturbative piece (coulomb or linear) should determine the perturbatively compatible effective radial separation between the quark and the anti-quark. \\

In the present work, we therefore show that only in the short distance range ($0\textless{r}\textless{r}^S$), linear potential is perturbatively compatible while for the alternative choice ( coulomb as perturbation), the corresponding range belongs to large distance ($r^L\textless{r}\textless{\infty}$). The exact magnitudes of $r^L$ and $r^S$ have explicit dependence on strong coupling constant $\alpha_{s}$, the confinement parameter $b$ and on the scale factor $c$. \\

The aim of the present paper is to outline the new features of the improved version of the model and present the prediction for the bounds on r.m.s. radii of pseudoscalar heavy flavored mesons (both heavy light and heavy heavy). Comparison is then made with those of other models available in literature.\\

In section 2, we outline the formalism, while in section 3 summarize the results and discussions. Section 4 contains conclusions.

\section{Formalism}

\subsection{Definition of r.m.s. radius}
The r.m.s. radius  \cite{12,13} of the bound state of quark and anti quark like meson is defined as
\begin{equation}
\langle r^2_{rms} \rangle=\int_0^{\infty}r^2 [\psi(r)]^2dr
\end{equation}
having radial wave function $\psi(r)$.

\subsection{Definition of charge radius}
The standard definition of charge radius \cite{14} is
\begin{equation}
\langle r^2_E \rangle=-6 \frac{d^2}{dQ^2}eF(Q^2)\vert _{Q^2=0}
\end{equation}
where $F(Q^2)$ is a meson form factor \cite{15,16}.

Using above equation, the charge radii of various heavy flavored mesons are calculated in ref.\cite{bjh,nsb}.

\subsection{Relation Between RMS and charge radius}
The charge radii can be measured with electromagnetic probe whether the RMS radii defined as in eqn.(1) cannot be measured with electromagnetic probe. The RMS radius is nearly the average $\textless{r}^2\textgreater$ of the quark wave function, which presumably may be determined in Quark Gluon Plasma (QGP) experiments presently studied at LHC \cite{17}. However a simple approximate relationship between the two can be found from the following equation
\begin{equation}
r_E^2=\sum_i e_i \left[ \textless{r_i}^2\textgreater + \frac{3}{4m_i^2} \int d^3 p \mid \Phi(p)\mid^2 \left( \frac{m_i}{E_i}\right) ^{2f}\right] 
\end{equation}
derived by Godfrey and Isgur \cite{18}. \\

Here $e_i$ is the charge of the $i^{th}$ quark/antiquark, $r_1$ and $r_2$ are the distances of the two quarks/antiquarks measured from the centre of mass, m is the mass of the quark and $E=\sqrt{p^2+m^2}$. $\Phi(p)$ is the quark momentum distribution. The exponent \textit{f} can be determined in a semi-empirical way.\\

We obtain the inequality
\begin{equation}
r_E^2 \textgreater \sum_i e_i  \textless{r_i}^2\textgreater
\end{equation} 

making the transformation of the co-ordinate we choose one of the quarks/antiquarks located at the origin. It results in 
\begin{equation}
r_E^2 \textgreater  e  \textless{r}^2\textgreater
\end{equation} 

where $\textless{r}^2\textgreater$ can be interpreted as the standard RMS as defined in eqn.(1) and $e=\sum e_i$.\\

It yields the desired results: the ratio of the charge radii and RMS radii of the meson should always be more than the square root of the charge of the quark/anti quark of the meson having the larger value. As an illustration for $D^+$ meson, the ratio should not be less than $\surd \frac{2}{3}$.

\subsection{ The total wave functions with an improved perturbative approach}

The Schrodinger equation describing the quark-anti quark bound state is
\begin{equation}
-\frac{\hbar^2}{2m}\nabla^2\psi(r)+(E-V)\psi(r)=0
\end{equation}

The standard QCD potential is defined as  \cite{14} 

\begin{equation}
V(r)=-\frac{4\alpha_s}{3r}+br+c
\end{equation}

Where -$\frac{4}{3}$ is due to the color factor, $\alpha_s$ is the strong coupling constant, r is the inter quark distance, b is the confinement parameter (phenomenologically, $b=0.183 GeV^2$  \cite{19}) and `c' is a constant scale factor.\\

For potential (7), we can make four choices:\\

Choice-I:$-\frac{4\alpha_s}{3r}$ as parent and br+c as perturbation\\

Choice-II:br as parent and $-\frac{4\alpha_s}{3r}+c$ as perturbation\\

Choice-III:$-\frac{4\alpha_s}{3r}+c$ as parent and br as perturbation and\\

Choice-IV:  br+c  as parent and $-\frac{4\alpha_s}{3r}$ as perturbation\\ 

It is well known in quantum mechanics that a constant term `c' in the potential can at best shift the energy scale, but should not perturb the wave function. This important point was overlooked in earlier publications  \cite{7,8,9} of the subject, but the present work takes into account this.\\

We use the Dalgarno's method of perturbation to construct the wave functions.\\

The wave function for choice-I \cite{7} including relativistic effect is

\begin{equation}
\psi^{total}(r)=\frac{N_1}{\sqrt{\pi a_0^3}}\left[ 1+cA_0\sqrt{\pi a_0^3}-\frac{1}{2}\mu b a_0r^2\right]\left( \frac{r}{a_0}\right) ^{-\epsilon} e^{-\frac{r}{a_0}},
\end{equation} 

where $A_0$ is the undetermined co-efficient appearing from the series solution of Schrodinger's equation and

\begin{equation}
a_0=\left( \frac{4}{3}\mu\alpha_s\right)^{-1}
\end{equation}

\begin{equation}
\mu=\frac{m_1m_2}{m_1+m_2}.
\end{equation}

Here $m_1$ and $m_2$ are the masses of quark and anti quark respectively and $\mu$ is the reduced mass of the mesons and

\begin{equation}
\epsilon=1-\sqrt{1-\left( \frac{4}{3}\alpha_s\right) ^2}
\end{equation}

provides the relativistic effect \cite{20,21} due to Dirac modification factor and $N_1$ is the normalization constant given by

\begin{equation}
N_1=\frac{1}{\left[ \int_0^\infty \frac{4 r^2}{a_0^3}\left[ 1+cA_0\sqrt{\pi a_0^3}-\frac{1}{2}\mu b a_0r^2\right]^2\left( \frac{r}{a_0}\right) ^{-2\epsilon} e^{-\frac{2r}{a_0}}dr\right] ^{\frac{1}{2}}}.
\end{equation}

Similarly, the wave function for choice-II upto $O(r^4)$ is calculated in ref. \cite{22} and is given by

\begin{equation}
\psi^{total}(r)=\frac{N_2}{r} \left[1+A_1(r)r+A_2(r)r^2+A_3(r)r^3+A_4(r)r^4\right]A_i[\rho_1 r+\rho_0] \left( \frac{r}{a_0}\right) ^{-\epsilon}
\end{equation}

where $A_i[r]$ is the Airy function \cite{23} and $N_2$ is the normalization constant
\begin{equation}
N_2= \frac{1}{\left[ \int_0^{r_0} 4 \pi \left[1+A_1(r)r+A_2(r)r^2+A_3(r)r^3+A_4(r)r^4\right]^2 \left( A_i[\rho_1 r+\rho_0]\right) ^2 \left( \frac{r}{a_0}\right) ^{-2\epsilon}dr\right]^{\frac{1}{2}} }.
\end{equation}

The co-efficients of the series solution as occured in Dalgarno's method of perturbation, are the functions of $\alpha_s, \mu, b$ and c:
\begin{equation}
A_0=0
\end{equation}
\begin{equation}
A_1=\frac{-2\mu \frac{4\alpha_s}{3}}{2\rho_1 k_1+\rho_1^2 k_2}
\end{equation}

\begin{equation}
A_2=\frac{-2\mu (W^1-c)}{2+4 \rho_1 k_1+ \rho_1^2 k_2}
\end{equation}

\begin{equation}
A_3=\frac{-2\mu W^0 A_1}{6+6 \rho_1 k_1+ \rho_1^2 k_2}
\end{equation}

\begin{equation}
A_4=\frac{-2\mu W^0 A_2+2\mu b A_1}{12+8 \rho_1 k_1+ \rho_1^2 k_2}
\end{equation}

The different parameters are given by,
\begin{equation}
\rho_1=(2\mu b)^{\frac{1}{3}}
\end{equation}

\begin{equation}
\rho_0=-\left[ \frac{3\pi (4n-1)}{8}\right] ^{\frac{2}{3}}
\end{equation}

(In our case n=1 for ground state)

\begin{equation}
k_1=1+\frac{k}{r}
\end{equation}

\begin{equation}
k=\frac{0.3550281-(0.2588194) \rho_0}{(0.2588194) \rho_1}
\end{equation}

\begin{equation}
k_2=\frac{k^2}{r^2}
\end{equation}

\begin{equation}
W^1=\int \psi^{(0)\star} H^{\prime} \psi^{(0)} d\tau
\end{equation}

\begin{equation}
W^0=\int \psi^{(0)\star} H_0 \psi^{(0)} d\tau.
\end{equation}

The above analysis illustrate that the scale factor `c' has observable effect in the final wave function if we apply Dalgarno's method of perturbation. Now in the present work we want to see whether the term `c' has also its affect even if we consider `c' in the parent part of the Hamiltonian.\\

Using Dalgarno's method of perturbation the total wave function for choice-III upto $O(r^4)$ is \\

\begin{equation}
\psi^{total}(r)=\frac{N_3}{\sqrt{\pi a_0^3}}\left[ 1-\frac{1}{2}\mu b a_0r^2-\frac{1}{20}\mu^2 bca_0r^4\right]\left( \frac{r}{a_0}\right) ^{-\epsilon} e^{-\frac{r}{a_0}}
\end{equation}

where the normalization constant $N_3$ is

\begin{equation}
N_3=\frac{1}{\left[ \int_0^{\infty} \frac{4 r^2}{a_0^3}\left[ 1-\frac{1}{2}\mu b a_0r^2-\frac{1}{20}\mu^2 bca_0r^4\right]^2\left( \frac{r}{a_0}\right) ^{-2\epsilon} e^{-\frac{2r}{a_0}}dr\right] ^{\frac{1}{2}}}
\end{equation}

Similiarly, the wave function for choice-IV including relativistic effect considering upto O($r^4$) is 

\begin{equation}
\psi^{total}(r)=\frac{N_4}{r} \left[1+A_1(r)r+A_2(r)r^2+A_3(r)r^3+A_4(r)r^4\right]A_i[\rho_1 r+\rho_0] \left( \frac{r}{a_0}\right) ^{-\epsilon}
\end{equation}

where $N_4$ is the normalization constant given by

\begin{equation}
N_4= \frac{1}{\left[ \int_{r^{long}}^\infty 4 \pi \left[1+A_1(r)r+A_2(r)r^2+A_3(r)r^3+A_4(r)r^4\right]^2 \left( A_i[\rho_1 r+\rho_0]\right) ^2 \left( \frac{r}{a_0}\right) ^{-2\epsilon}dr\right]^{\frac{1}{2}} }.
\end{equation}

The co-efficients are

\begin{equation}
A_0=0
\end{equation}

\begin{equation}
A_1=\frac{-2\mu \frac{4\alpha_s}{3}}{2\rho_1 k_1+\rho_1^2 k_2}
\end{equation}

\begin{equation}
A_2=\frac{-2\mu W^1}{2+4 \rho_1 k_1+ \rho_1^2 k_2}
\end{equation}

\begin{equation}
A_3=\frac{-2\mu (W^0-c)A_1}{6+6 \rho_1 k_1+ \rho_1^2 k_2}
\end{equation}

\begin{equation}
A_4=\frac{-2\mu (W^0-c)A_2+2\mu b A_1}{12+8 \rho_1 k_1+ \rho_1^2 k_2}.
\end{equation}

The equations (27) and (29) showed that even if we consider the scale factor `c' in the parent Hamiltonian, then also it affects the total wave function.\\

Therefore, Dalgarno's method in general conflicts with the quantum mechanical notion that the scale factor `c' should not have an observable effect except the energy shift.\\

There are two possible ways of making the perturbative approach adequate to quantum mechanics. At phenomenological level one keeps the scale factor in the potential so that it can contribute to the short and long range of the inter quark distance without changing the total wave function in a true sense. A more theoretically satisfactory approach will be to limit the domain of the applicability of Dalgarno's method, that it is not applicable to potential having non-zero scale factor. This approach is then corrected quantum mechanically. We take this latter view point in the present paper.\\

For the validation of the quantum mechanical expectation we therefore consider $c=0$ in the potential (7). In such situation choice-I $\&$ II and III $\&$ IV are identical.\\

Now to find the cut offs we use perturbation conditions:\\

Case-I: For coulomb as parent and linear as perturbation:
\begin{equation}
-\frac{4\alpha_s}{3r} \textgreater br
\end{equation}

and\\

Case-II: For linear as parent and coulomb as perturbation:

\begin{equation}
br \textgreater -\frac{4\alpha_s}{3r}
\end{equation}

From (36) and (37) we can find the bounds on r upto which case-I and II are valid. Case-I gives the cut off on the short distance $r_{max}^S<\sqrt{\frac{4\alpha_s}{3b}}$ and case-II gives the cut off on the long distance $r_{min}^L>\sqrt{\frac{4\alpha_s}{3b}}$.\\

In table-1, we show the bounds on $r^S$ and $r^L$ in fermi which yields exact/most restrictive upper bounds of the quantities to be calculated.

\begin{table}[h]
\caption{$r^S$ and $r^L$ in fermi with b=0.183 $GeV^2$}
\begin{center}
\begin{tabular}{|l|l|}
\hline
            Scale            & $r^S=r^L$ (fermi)\\ \hline
                      
            c-scale          &    \\
    ($\alpha_s=0.39)$        & 0.33207\\ \hline
                     
             b-scale         &    \\
   ($\alpha_s=0.22)$         & 0.24941\\ \hline
\end{tabular}
\end{center}
\end{table}                     

Again the application of Airy function as meson wave function needs suitable cut off to make the analysis normalizable and convergent. We therefore set the cut off ($r_0$) in the range 1 fm (5.076 $GeV^{-1}$) \cite{24} for our calculations.\\

The equation (1) of section 2.1 is thus modified to

\begin{equation}
 \textless{r_{rms}}^2\textgreater= \int_0^{r^S}r^2 [\psi(r)]^2dr + \int_{r^L}^{r_0}r^2 [\psi(r)]^2dr
\end{equation}

\section{Results and discussion}

Using equation (38) and with the help of the wave functions from equations (27) and (29) we compute the bounds on RMS radii of various mesons in the ground state $(l=0)$ and results are tabulated in table-2. However it is to be noted that $r^S$ and $r^L$ are the perturbative saturation lengths for coulomb parent and linear parent respectively. The proper perturbative range should be far less than the cut offs $r^S$ and $r^L$. The upper bound corresponds to the sum of the contribution of the two options at short (coulomb parent) and long distances (linear parent), while the lower bound correspond to the minimum of the two (linear parent).\\

The input parameters in the numerical calculations used are same as our previous work  \cite{25,kkpijmpa,kkpnsb}; $m_u=0.336 GeV$, $m_d=0.336 GeV$, $m_s=0.483 GeV$, $m_c=1.55 GeV$ and $m_b=4.95 GeV$, b=0.183 $GeV^2$.
\newpage

\begin{table}[h]
\caption{Bounds on RMS radii ($r_{rms}$) in $fm$ with c=0}
\begin{center}
\begin{tabular}{|l|l|l|}
\hline
$\alpha_s$                   & Meson          & Bounds on   \\ 

& &  ($r_{rms}$) in $fm$ \\ \hline

\multirow{5}{*}{0.39}  & $\pi^+(u\bar{d}$) &  0.2914-1.1334 \\
& $K(u\bar{s}/d\bar{s})$ & 0.2940-1.1184\\
& $D(c\bar{u}/c\bar{d}$) & 0.3008-1.0924\\ 
& $D_s^+(c\bar{s})$ & 0.3091-1.0747\\ 
&J/$\psi(c\bar{c})$ & 0.3489-1.0530\\ \cline{1-3} 
  
\multirow{2}{*}{0.22} &  $B_c^+(\bar{b}c)$  & 0.3841-1.0942 \\
& $\tau(b\bar{b})$ & 0.4439-1.1331 \\ \hline
 
\end{tabular}
\end{center}
\end{table}

In table-3, we give the different model predictions of rms radii for heavy flavored mesons available in literature  \cite{26,27,28}.
\begin{table}[h]
\caption{$r_{rms}$ values in $fermi$}
\begin{center}
\begin{tabular}{|l|l|l|l|}
\hline
\multirow{2}{*}{Meson} & \multicolumn{3}{l|}{\ \ \ \ \ \ \ $r_{rms} (fm$)} \\ \cline{2-4} 
                       &    Ref.[30]   & Ref.[31]  & Ref.[32] \\ \hline
 $c\bar{c}$            & 0.4490        & 0.4453    & 0.4839 \\ \hline
 $b\bar{b}$            & 0.2249        & 0.2211    & 0.2672 \\ \hline
\end{tabular}
\end{center}
\end{table}

From table-3, we see that the value of rms radii for $c\bar{c}$ is well within the range of table-2, but the result for $b\bar{b}$ is out of range. A higher value of cut off parameter $(r_0)$ in this sense can improve the results for $b\bar{b}$ and then the other theoretical predictions can be accomodated within the range.\\

In this work we have taken the cut off $r_0=$1fm (5.076 $GeV^{-1}$) \cite{24}. However from the result we can conclude that the scale for light-light, heavy-light and heavy-heavy mesons should not be same, rather it should be increased with the increase of the mass scale in a particular way.\\

From table-2, the bounds on $r_{rms}$ for $D^+(c\bar{d}$) is 0.3008-1.0924 and hence the corresponding value of charge radii $(r_E)$ should be more than this bound as total charge of the meson is +1 and this fact is found to be within the prediction. Thus the validitity of equation (5) is also tested.

\section{Conclusions}
We have applied the perturbatively compatible approach to the QCD potential model studied earlier \cite{7,8,9,22,25} and computed the upper and lower bounds of the RMS radii of heavy light mesons. The theoretical prediction of bound for $c\bar{c}$ meson is shown in table-3 . However the `c' itself cannot be determined from the perturbative analysis. There is a suggestion in ref. \cite{7} that the scale should not exceed $\sim$ 1GeV for heavy flavored mesons which is consistent with models of ref. \cite{29,30,26,27,28}.\\

Let us now comment on the physical significance of RMS radii of heavy flavored mesons and their plausible experimentally observable possibilities. The present work will hopefully be useful to extract information on RMS radii from the charge radii itself for heavy flavored mesons, where model predictions are available in current literature \cite{31,32,bjh}. We also check the sensitivity of the cut off parameter $(r_0)$ which is taken to be $1fm$ in this work and the results show that the cut off parameter should not be same for light-light, heavy-light and heavy-heavy mesons, rather the cut off parameter should be increased from light to heavy mesons. Moreover it is to be noted that the coupling constant ($\alpha_s$) is chosen only in two scales: charmonium $(\alpha_s=0.39)$ and bottomonium $(\alpha_s=0.22)$ scales. A scale dependent $\alpha_s$ \cite{faustov} might be useful in this context.\\

Although several theoretical models are available in literature, but there is no present experimental information either on charge radii or RMS radii of heavy flavored mesons. Hopefully, future experiments eg., PANDA \cite{34}, LHCb  \cite{17} will yield more information to test the validity of the simple relations as well as data.

\section*{Acknowledgement}
\textit{One of the authors (TD) acknowledges the support of University Grants Commission in terms of fellowship under BSR scheme to pursue research work at Gauhati University, Department of Physics.}

\end{document}